# Exploring Cultures through Pattern Mining
# - Practices from Generative Beauty Workshops

Jei-Hee Hong[1], Yuma Akado[1], Sakurako Kogure[1],
Alice Sasabe[1], Keishi Saruwatari[2] and Takashi Iba[1]

[1]Keio University
Endo 5322, Fujisawa, Kanagawa, 252-0816, Japan
[2]Kao Corporation
Kotobukicho 5-3-28 Odawara, Kanagawa, 250-0002, Japan
e-mail: ilab-beauty@sfc.keio.ac.jp

**ABSTRACT**

This paper presents a method for understanding personal ways of thinking and doing in daily lives among different countries by mining their ways as patterns in a sense of *pattern language*. Pattern language is a methodology of describing tacit practical knowledge, where each pattern consists of context, problem, and solution. In this paper, patterns mined from the workshops we held in the following three countries: Japan, Korea, and the United States, are analysed. The results demonstrate similarities and reflect characteristics of the patterns of each country. We anticipate that this workshop can be used as a method for better understanding of cultural similarities and features in the light of practical knowledge in daily lives.

**INTRODUCTION**

People have their own tips for living that can be used in their daily lives. It is also the same with the way, how women live lively and beautifully. Those tips could be noticed by one person, but not found by others. Therefore, we made a pattern language called *Generative Beauty Patterns* to make it possible to share hints and tips for living beautifully and lively (Arao, *et al.*, 2012).

The Generative Beauty Patterns were mined mainly for Japanese people to support their daily lives in Japan. These patterns were translated into English in order to present in other countries. We received feedbacks saying that these patterns contain typical Japanese cultural aspects. These comments encouraged us to get more interested in similar or unique characteristics of cultures when we minded patterns for tips or hints in daily lives. Thus, we designed and hosted pattern mining workshops in three countries.

In this paper, we will propose the results of the pattern mining workshops we held in Japan, Korea and the United States. Moreover, we will also discuss about the possibility of cultural understanding that can be implied from the results of workshops.

**WORKSHOP FOR MINING PATTERNS**

In the workshop, practical knowledge on "How to live their beautiful life to the fullest" is mined from dialogues among the participants. The participants share their tips or concerns with others. Through this way, necessary data for writing patterns can be mined: their tips can be a *solution* and their concerns can be a *problem* for a pattern.

We held such workshops for students and businesswomen for five times in Fujisawa and Tokyo, Japan, 2014; for Korean students from eight universities in Seoul, Korea, 2014; and for students and faculty members at the University of North Carolina Asheville, USA, 2014. The total number of participants in each country is 70 in Japan, 8 in Korea, and 25 in USA (Figure 1).

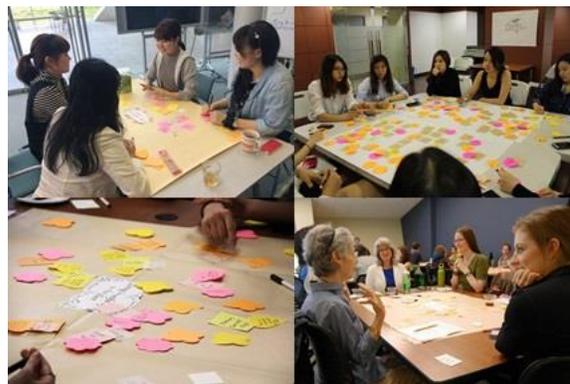

*Figure 1. Pattern Mining Workshops We Held in Japan, Korea and the United States.*

**OBTAINED PATTERNS**

In this section, we present the patterns mined from our workshops. Note that these patterns do not completely represent each culture, as the condition of workshop was not fully controlled. Therefore, the patterns introduced here are only the examples, not the crucial representatives.

**Patterns Mined in Each Country**

By holding workshops, we obtained 290 seeds of patterns in Japan, 40 in Korea, and 147 in the USA. In what follows, we introduce some interesting patterns from the three countries.

The followings are examples of Japanese patterns: "Own Style" tells to own their style by keep wearing it on; "Self Reward" suggests people to prepare treats for themselves after work; and "Appropriate Distance" advices to put themselves to the first priority by keeping a reasonable distance with others.

Patterns from Korea are "My People", which suggests not to keep concerns or feelings but to share it with people they trust; and "Size Guide", which advices people to know their body size to make online shopping more effective.

There are three examples of the patterns that are mined from the USA. "Comfy Choice" tells people to choose the clothes they feel comfortable with, "Eye Match" suggests people to coordinate items that matches their eye colours, and "Positive Self-Talk" recommends people to talk positively to themselves when feeling down.

**The Similarities among the Countries**

By comparing the patterns mined from each country, we found that there are some similarities. These patterns might be considered as important factors for living beautiful life in every country.

The first example is "NO-ing expert" from Japan, and "Courage to say NO" from the USA. "NO-ing expert" encourages people to be brave to cut off schedules with low priority to earn time for themselves. "Courage to say NO" simply suggests to be brave enough to say no if it is necessary. These patterns are similar in a way in which both tell the importance of putting themselves as a priority by saying "no".

The second example is "The First Step" from Korea and "From Myself" from Japan. The main theme of the two patterns is how to generate an interactive conversation when meeting people for the first time. "The First Step" implies people to take the lead in the conversation. These two are similar as "From Myself" also advices people to talk about themselves first.

**The Characteristics of the Countries**

We found that some patterns also reflect the characteristics of each country.

"Eye Match" mined in the USA reflects USA's diversity. This pattern recommends people to coordinate items that match their eye colours. This pattern might not have been mined from Japanese nor Korean as most of them have dark coloured eyes. There is no need for them to think about their eye colours when choosing their clothes. Therefore, "Eye Match" reflects the USA or Western culture, where people have various eye colors.

"Self-Reward" from Japan and "Positive self-talk" from the USA show different approaches on how to encourage themselves. "Self-Reward" is about encouraging self by preparing treats after a hard work. "Positive Self-talk" suggest people to think of positive aspects of them in order to encourage themselves. These two patterns are different as the source of the motivation comes from the surroundings in the former pattern, while the latter has it within a person.

Another type of features we found is the difference in action. This means that they have similar viewpoint but different implementations. The example is the pattern no. 31 of Generative Beauty Patterns, "*Cheer-up Cookie*". It suggests people to know how to cheer them up. The actions mined in Japan are to listen to the music or eat delicious food; on the other hand, the action mined in Korea is to eat spicy food. As Korean people like to eat spicy food, they sometimes choose this method to release their stress. Like this, we could learn that actions can vary as it reflects the culture of the country even on the same pattern.

**CONCLUSION**

In this paper we presented a method to understand personal ways of thinking and doing in daily lives among different countries. The results of workshops in three countries showed similarities and characteristics of each culture. This analysis consists of ordinary people's thoughts or habits in their daily lives. Although the number of patterns we mined so far is not enough to make a concrete statement, we think our study can be used as a method for the research on cross-cultural understanding as we could analyse from the dialogue of participants.

**ACKNOWLEDGMENT**

We would like to thank Prof. Mary Lynn Manns for supporting us on holding workshop at University of North Carolina at Ashville.